[1994/12/01]% LaTeX date must December 1994 or later
\documentclass[draft]{article}
\title{Genetic Algorithms and Critical Phenomena}

\author{   A.  Barra\~n\'on  
\footnote{ Universidad Autonoma Metropolitana. Unidad Azcapotzalco.
Av. San Pablo 124, Col. Reynosa-Tamaulipas, Cd. de M\'exico. 
email: bca@correo.azc.uam.mx } ;
 J. A.  L\'opez
\footnote{Dept. of Physics, The University of Texas at El Paso. El Paso, TX, 79968  }    ;
C. Dorso.
\footnote{ Depto.de F\'{\i}sica, Universidad de Buenos Aires. Buenos Aires, Argentina}    }

\date{Nov $20^{th}$ 2002}

\usepackage{amsmath,amsthm}

\chardef\bslash=`\\ % p. 424, TeXbook
\newcommand{\ntt}{\normalfont\ttfamily}
\theoremstyle{definition}

\theoremstyle{remark}

\newcommand{\eval}[2][\right]{\relax
  \ifx#1\right\relax \left.\fi#2#1\rvert}

\newcommand{\envert}[1]{\left\lvert#1\right\rvert}

\begin{document}
\maketitle
\markboth{Sample paper for the {\protect\ntt\lowercase{amsmath}} package}
{Sample paper for the {\protect\ntt\lowercase{amsmath}} package}
\renewcommand{\sectionmark}[1]{}

\abstract

Genetic algorithms based on natural selection and minimal fluctuations 
have been applied to model physical and biological systems. Critical exponents 
have been extracted via computational simulations of nucleation for colossal 
magnetoresistance, heavy ions liquid-gas phase transitions and HIV to AIDS transition.  

%\keyword{critical phenomena, heavy-ion reactions,percolation,
%multi-fragmentation \PACS{$ 25.70.Pq$}

\section{Introduction}
We have developed genetic algorithms to study the evolution of several 
finite physical systems starting from an initial state, until an absorbing state 
is reached. Phase transition is identified with fragment build up and the critical 
multiplicity is the one providing the best $\chi^2$ test.

Since Fisher Liquid Droplet Model  as well as Random percolation show a power
 law near the critical point,  we may speculate whether both belong to the same 
universality class. In both cases finite size and geometry effects are also anticipated. 
Even when these systems have a different physical nature, the fact that correlations 
are increased close to the critical point leads us to question whether these 
belong to the same universality class. This is extended to the VIH to AIDS transition, 
which can be modeled as a nucleation process whose critical exponents were 
hereby extracted.

The behavior close to the critical point is characterized by a loose of time and
space scales, which promotes the similarity among all the critical 
phenomena, independently of their specifics. In the case of the transient and finite 
systems, the most important scales are system size and reaction duration, though
 critical phenomena will not always occur. Hence, it is important to prove the
 feasibility of critical behavior in these finite dynamical systems.

First neighbor particle correlations are not
 present  in gases though liquids exhibit strong first neighbor two body
 correlations. Hence, correlations grow  close to the critical point,
including all particles of the system. Therefore, 
the influence of geometry should be considered as well as class universality  
of different physical system should be questioned \cite{Onsager}. 

Even when power laws have been observed in distinct physical systems, such as
 seismic distributions, fractal coastlines, weather records and DNA statistics
\cite{Grass}, it is well known that log-log plots are able to depict spurious 
linear relations. This would result from a self-organized criticality ($SOC$), where 
non equilibrium systems should move slowly through an instability with a global 
and local relaxing mechanism. Also, these anomalous power laws can be related to 
codimension one critical points, where a critical behavior is reached after a 
temperature fluctuation even when this could not arise from a natural evolution 
of the system \cite{Grassber}. Therefore, it is important to perform simulations of 
theoretical models in order to prove the feasibility of the critical 
behavior associated with these power laws.

Cohen et. al. extracted of critical exponents scale-free networks, where 
a power law holds: $P(k)=ck^{- \lambda}, k \ge m$, with 
$P(k)$ equal to the probability of a node being connected to exactly $k$ nodes
 \cite{Cohen}. Grassberger obtained percolation thresholds for simple
 hypercubic lattices whose dimension goes from 4 to 13, finding anomalous
 scaling when $d \ge 7$: $M(t) = M_{\infty} - const. /t^{\omega}$, where $M(t)$ 
is the number of wetted sites at step $t$
and the scale correction is in agreement with the following power law:
$M_{\infty}-1 \sim (d-6)^{- \alpha}$ \cite{Grassber1}. 

Fortunato and Satz studied the explicit symmetry breakup for spin models, 
finding that in the case of a second order phase transition with a small field, the 
percolation thresholds line lies parallel to the susceptibility peak line 
(pseudo critical line). And the first order phase transition line is a 
percolation transition line, though the critical exponents are not equal to those 
belonging either to the random percolation model or the Ising model
 \cite{Fortunato}. This leads to consider a new cluster definition that takes on
 account the magnetic field intensity, which could be used to study QCD
 deconfinement with dynamical quarks \cite{Satz}.

   Coniglio postulated that above the critical dimension, the number of percolation 
clusters is infinite and hyperscaling relation no longer holds :
 $2-\alpha=\nu d$ \cite{Conig1}. 
This was proven by Andronico et. al. for the case of random percolation, 
though this does not hold for the Ising model due to distinct behaviors 
below and above the critical temperature \cite{Andronico}. 

The manuscript is organized in the following way. In the second section Fisher nucleation 
theory is described, showing how to extract the critical exponents close to the 
critical multiplicity. In the third section some reasons are given to expect 
this power law for a percolation system and the expected values for some percolation 
systems are shown. In section IV a genetic algorithm for HIV to AIDS transition is described, 
and its critical exponent is extracted. In section V a genetic algorithm is 
used to compute the critical exponent of the Colossal Magnetic Resistance. Section VI 
describes the Molecular Dynamics simulation employed to compute the critical exponent
 for the liquid-gas phase transition in Heavy Ions as well as the simulated annealing
 algorithm used to detect clusters. In section VII some conclusions are established 
from these results.

\section{Fisher Liquid Droplet Model}\label{NT}

According to fluctuation theory~\cite{landau}, the probability to obtain a 
liquid droplet of radius $r$ and $A$ nucleons in a vapor at temperature $T$
is given by~\cite{lopezdorso}:
\begin{eqnarray*}
P_{r}(A)=Y_{0}A^{-\tau }e^{-\left[ \left( \mu _{l}-\mu _{g}\right)
A+4\pi r^2_{0}\sigma (T)A^{2/3}\right] /T}  \ ,
\end{eqnarray*}
where $\Delta G$ is the change due to the phase transition in Gibss Free energy.  
This includes surface, curvature and bulk energy terms.
In the coexistence region, nevertheless, 
  $\left( \mu _l-\mu_g\right) =0$, and:
\begin{eqnarray*}
P_r(A)=Y_0 A^{-\tau}\exp [{-4\pi
r_0^2\sigma(T)A^{2/3}/T}]
\end{eqnarray*} 
Finally, in the critical point:
$\left( \mu _l-\mu _g\right) =0$, 
but as the liquid and vapor are indistinguishable at this point, surface energy 
is null:
$\sigma (T_c)=0$, 
hence the distribution is
\begin{eqnarray}\label{powerl}
P_r(A)=Y_0A^{-\tau }  \ ,
\end{eqnarray}
namely a pure power law, and as such, scale free. The exponent
 $\tau$ appearing in the droplet size distribution~(\ref{powerl}), 
is known as the critical exponent since it is a dimensionless constant with
 a common value for different systems.

Fisher Liquid Droplet Model ($FDM$) for nucleation~\cite{fisher} refines the probability
~(\ref{powerl}) to obtain a critical mass distribution normalized to the size of the 
system:
\begin{equation}\label{nA}
n_{A} =q_{o}A^{-\tau}
\end{equation}
with a proportionality constant $q_{o}$ that can be obtained using the 
first moment, $M_{1}=\sum_{A}n_{A}A$, of the normalized mass distribution,
 ({\it i.e.} $M_{1}=1$). Therefore 
 $q_{o}$ 
can be obtained with the following equation:
$q_{o}={1}/{\sum_{A}A^{(1-\tau)}}$. In order to select critical events, 
critical multiplicity is identified as the one providing the optimal
$\chi^2$ fit~\cite{dorlop2001}.

Supposing scaling for the fragment size distributions, close to the 
critical point:
\begin{equation}  M_k \sim   \envert{ T - T_c  }^{ (-1- k + \tau )/ \sigma} ,
\end{equation}
 where: $ 2 < \tau < 3$.
   Critical exponents are given by the following expressions   
 \cite{Gil1}:
\begin{equation}
M_2 \sim { \envert{ \epsilon}}^{- \gamma } 
\end{equation}
\begin{equation} A_{max} \sim { \envert{ \epsilon}}^{ \beta } 
\end{equation}
when $m=m_c$. 

\section{Percolation critical exponent}

A great variety of percolation methods have been used ever since Flory
 introduced percolation in the context of polymer gellation \cite{Flory}.
For many spin models, there exists a mapping of the equivalent graphic 
representation of a percolation transition corresponding to a spin model 
phase transition \cite{Chayes}. 

  Harreis and Bauer introduced a method to deal with N component percolation, 
finding new first order phase transitions and new empirical formulas for the 
percolation threshold as a function of component concentration \cite{Harreis}.
 Bauer introduced percolation in the study of fragmentation~\cite{bauer_2,bauer3},
 vid. Stauffer~\cite{stauffer} for more details. 

A percolation cluster has activated bonds going from one side of the 
lattice to the opposite side. For infinite systems, there is a well defined 
``critical probability" $p_{c}$, above which the probability to find a 
percolation cluster is equal to $1$, meanwhile bellow $p_{c}$ this probability is
 equal to $0$. For finite lattices, this transition is soft, {\it i.e.} 
the probability to find a percolation cluster is not equal to $0$
for any probability. Mader ${\it et. al.}$ 
have shown reducibility and thermal scaling properties in the Ising model, 
obtaining a value of $\tau = 2.39$ \cite{Mader}. Chayes et.al. have 
shown a finite cluster scaling critical exponent $\tau -2=1/2$
and an infinite cluster scaling critical exponent $\tau -2=1/2$
\cite{Chayes1}. 
In the percolation model the weight of a given configuration
 $C$ of n links is given by:

\begin{equation}
W(C)=p^n(1-p)^{N-n}
\end{equation}
where N is the number of vertices in the lattice. Close to the percolation 
threshold, the critical behavior is characterized by the following critical 
exponents:
\begin{equation}
P_{\infty}=1- \sum s n(s,p) \sim \envert{p-p_C}^{\beta}
\end{equation}
and:
\begin{equation}
S(P)=\sum s^2 n(s,p) \sim \envert{p-p_C}^{\gamma}
\end{equation}

Cluster distribution satisfies the following scaling relation:
\begin{equation}
n(s,p)= s^{- \tau} f( (p-p_C) s^{\sigma} )
\end{equation}
hence a power law is expected close to the critical point:
\begin{equation}
n(s,p)= s^{- \tau} f(0)
\end{equation}

Starting from these relations, the following equations can be obtained :
\begin{equation}
\tau= 2+ \frac{\beta}{\beta + \gamma}
\end{equation}
and:
\begin{equation}
\sigma= \frac{1}{\beta + \gamma}
\end{equation}
In 3D, the best estimation is $\tau = 2.18$ and $\sigma= 0.45$ \cite{Coniglio}

Bauer y Golinelli have obtained a second order phase transition when
 $\alpha = e=2.718...$ in the percolation of random graphs whose connectivity is 
 $\alpha$ and where the leaves and their neighbors are iteratively removed.
In that study, a new power law was found for $N_C$, the number of vertices in 
the core tree: $N_C \sim N^{\omega}$, where 
$N$ is the number of vertices of the original graph \cite{MBauer}. 
Power laws have been found in statistical distributions of Hamming distances 
for random threshold networks and random binary lattices with scale-free 
distributions whose interior degree $\sim k_{in}^{- \alpha}$  \cite{Rohlf}.

Kamp and Bornholdt have studied percolation transition in a directed graph, with 
activating and deactivating bonds, and extracted power law critical exponents for 
the number of avalanches as a function of active sites, as well as a plot of the 
fraction of activating bonds versus system size \cite{Kamp}. 

Bornholdt and Rohlf have studied a symmetrically connected threshold lattice whose 
topology is changed with a local rule, obtaining a power law in the plot of 
mean connectivity versus system size \cite{Bornholdt}.    

\section{HIV Cellular Automaton}

 Ebel et. al. found power laws for the avalanche size distribution 
 given by the number of mutations required to establish an equilibrium in
evolutionary games with Nash equilibriums perturbed by mutant events \cite{Ebel}.

When mutation rates are close to a substitution for each genome of each generation, 
a virus population forms a highly diversified cloud of mutants \cite{Burch}. 
Since sequence space is quite large, even for a population size of the order of 
$10^{12}$ there is a constant flux of brand new mutants.
With some probability these mutants will fixate, where fixation is understood as 
the moment in which a mutant comes out to be the ancestor of a new quasi-species 
that fully replaces the old one.

In the quasi-species configuration, on the other side, most of the progenies of 
a mutant suffers subsequent mutations and also its progenies. Therefore, a new 
species adaptation is not given by the adaptation of the first mutant, but by the 
the mean adaptation of the eventually formed mutant cloud.

´Wilke has shown, using a multibranching processes model, that fixation depends 
on the global rate of growth of the quasi-species  obtained due to an 
infection \cite{Wilke}. Asexual organisms genomic mutation rates, such as bacteria and
ADN, have been observed in the range of $2$x$10^{-3}\sim$$4$x$10^{-3}$,
hence a few in a thousand mutate \cite{Drake}. Smaller genomic mutation rates have 
not been observed since adaptability to a changing environment is required for a 
species \cite{Nilson}. Wilke has proved, based on quasi-species model, that in the 
case of low mutation rates, population could benefit on environment fluctuations
\cite{Wilke1}.  

\subsection{HIV Cellular Automaton}

Population evolution has been modeled using Kamp cellular automaton, reproducing
 HIV to AIDS transition, though in this study we focus on the 
nucleation of infected cells as well as the seldom observed incapability of HIV to 
evolve into AIDS \cite{Kamp2}. According to Kamp, in an infection, viral genomes 
are diversified due to mutation and selective pressure of the immune system. This 
coevolutionary dynamics can be modeled in sequence space. Viral genomes can be 
represented by chains of length $n$, built up from an alphabet of length 
$\lambda$ and their diversification can be described as a dispersion in sequence 
space. Analogously, we assign a sequence to the immune receptor corresponding to 
the viral strain. Any chain in sequence space is supposed to represent a viral 
epitome, as well as its complementary immune receptor.

Therefore, each sequence is characterized by a viral and immune 
state variable. A site in sequence space is susceptible if it can host a virus. 
It is called infected if the system has a virus with an epitome motif represented 
by the chain of the site. If a viral sequence finds an immune response it is removed 
and the system is immunized against it. In this case, and in the case where a site 
is unreachable for virus, it is called recovered (or removed).

Since viral and immune entities replication is affected by copy fidelity:
 $q_{vs} < 1$ and $q_{is} < 1$, system shows viral and immune dispersion 
in sequence space. Introducing some viral strains in a system originally free of them, 
a dynamic evolution follows from the cellular automaton approximation after 
the following iteration steps \cite{Kamp2}:

Kamp automaton proceeds in the following way. A random site is chosen and a) 
if the site represents an immune receptor, a randomly chosen bit is mutated with
 probability (1-$q_{is}$), or, b) if a new immune strain is generated and the mutant
 coincides with a site, the site is considered as a recovered site. If the site
 is infected, a bit is randomly changed with probability (1-$q_{vs}$), or, if a
 new immune strain is generated and the mutant coincides with a susceptible site,
 the site is considered as infected.

\subsection{Results}

A critical exponent was extracted with a value of $\tau = 2.32$ for the power law 
of infected cells of HIV to AIDS transition, simulated with a genetic algorithm 
that evolves populations with a cellular automaton (Fig. 1). 

Besides, infected site ratios were obtained as a function of genetic variability 
of both infected sites $q_{vs}$ and healthy sites $q_{is}$. As can be appreciated in 
(Fig. 2), two regions of low infected sites ratios are formed,
which ensures an immune response, for low and high values of  
genetic variability. This explains the observed capability of some immune 
systems to resist viral infections, notwithstanding high genetic mutation rates would 
lead to the opposite scenario.  
     
\section{Colossal Magnetoresistance}\label{cmr}

L\"ubeck has computed the critical exponents and the fluctuations of the order 
parameter, for the case of a conservative lattice, finding a maximum critical 
dimension for this gas equal to 4 \cite{Lubeck}. Mari ${\it et. al.}$ 
have performed corrections for the finite size of the system, in the case of the 
Binder parameter of the 3D binomial Ising spin glass \cite{Mari}. 
Janssen ${\it et. al.}$ have shown that for a vector magnetic system of order N, 
when the temperature is much greater than the critical temperature, and 
the system is suddenly compressed until the critical state, a dynamical scaling 
is installed in the early evolution of the system \cite{Janssen}.           

Ying ${\it et. al.}$ found a relation between the binding randomness and the 
critical universality for the Potts random binding ferromagnet with a trinary 
distribution of compressed disorders in triangular lattices \cite{Ying}.

Sim\~oes ${\it et. al.}$ studied the early evolution dynamics for the
 two dimensional Ising model with three spin interactions in a direction,
 taking on account the symmetry of the Hamiltonian and the boundary conditions
 when computing magnetization. They obtained the same critical exponents of
 the four states Potts model \cite{Simoes}.

Ying ${\it et. al.}$ studied the early dynamics and critical universality for 
the Potts model with q=2 and q=3 in triangular two dimensional lattices, 
obtaining critical exponents equal to those of the corresponding 
two dimensional square lattice, concluding that they belong to the same
 universality class \cite{Ying1}.

 Acharyya et. al. computed the critical temperature of the metamagnet 
$FeBr_2$, using the anisotropic classical Heisenberg model,
in a tetragonal lattice, finding that for high temperatures there is a
first order phase transition between a paramagnetic phase and a flipped spin 
tilted phase. And for low temperatures, the phase transition is
 discontinuous and from a flipped spin tilted phase up to a longitudinal
 ferromagnetic phase \cite{Acharyya}.

In this study three dimensional configurations were generated, randomly
 assigning three spin values $S=-1,0,1$, with the following Hamiltonian:

\begin{equation}
H= -J \sum_{i,j} S_i S_j - H \sum_i S_i .
\end{equation}
 
A site is randomly chosen and its value is changed with the following 
probability:
\begin{equation}
p=\frac{ e^{- \Delta H/T}}{1+ e^{- \Delta H/T}}
\end{equation}

\subsection{Results}

The extracted critical exponent for the colossal magnetoresistance is
 $2.39$ (Fig. 3). Besides, we have computed a critical exponent
 $\beta= 0.38$, suggesting that it belongs to the universality class of 
Heavy Ion Collisions, as shown in Table 1. The obtained value of $\beta$ is close
 to the one reported by Kudzia ${\it et. al.}$ in fragmentation experiments on Au
 emulsions \cite{Kudzia}. The spontaneous susceptibility plot 
 indicates a critical temperature close to 4.5 K (Fig. 4). 

\section{Simulations of Heavy Ion Collisions}\label{mol}

Among the signals used to search for nucleation in nuclear systems, the 
presence of a peak in the specific heat has been used as a phase 
transition signature for periodical systems \cite{Dorso5}. 

Another signature explored is the controversial power law of the fragment 
distribution, close to the critical point of a liquid-gas phase transition
 \cite{Mast1}. In the energy range of relativistic Heavy Ion Collisions, 
a phase transition has been associated with the onset of shock wave front
 instabilities  \cite{Bar1} .

In Heavy Ion Collisions, a phase transition is expected to turn into a 
second order phase transition in the critical point of the phase diagram (Fig. 5).  
Experimental measurements of Au+Au collisions performed at GSI, were 
used to build a plot of temperature versus excitation, providing evidence of phase
 coexistence, in agreement with predictions of statistical 
multifragmentation models excluding volume \cite{Gross1}. 
Other experiments performed in Bevalac extracted critical exponents and studied 
the dependency of both the second moment of charge distribution and the biggest 
fragment size on charged particles multiplicity. These data were 
consistent with a second order phase transition predicted by the 
percolation model \cite{Bauer1}. 

\subsection{Molecular Dynamics}

A Molecular Dynamics realistic model, nicknamed 
 ``LATINO Model''~\cite{oaxt1999,oaxt2001} has been used.
This three dimensional model uses a Pandhripande binary potential that reproduces
 the energy and empirical density of nuclear matter, as well as realistic effective
 scattering cross sections. Nuclei used hereby were spherical droplets with 
the desired number of protons and neutrons, produced as ground states using 
Molecular Dynamics. Once a spherical nuclear system is randomly created at a high temperature, 
 it is cooled until it reaches a self-contained state. At this moment, the 
confining potential is removed and the system is cooled until it reaches a 
reasonable binding energy. Collision is simulated boosting one of the nuclei against 
the other and integrating the coupled equations of motion using a Verlet algorithm, 
ensuring an energy conservation better than a $0.01\%$ (Fig. 6). 

\subsection{Fragment Recognition}\label{FR}

In order to transform particle information provided by Molecular Dynamics 
in terms of fragment information, an Early Cluster Recognition Algorithm is 
needed, such as the algorithm that finds the most bound partition of the system
~\cite{dorso93}, {\it i.e.} the set of clusters $\{ C_i \}$ for which the sum 
of internal energies of the fragments attains a minimum value:
\begin{eqnarray}
{ \{C_i\} }   & { = \atop {}} &  { \hbox{argmin} \atop {\scriptstyle \{C_i\}}  } { \textstyle {[E_{ \{C_i\}} = \sum_i E_{int}^{C_i}]} \atop {} } \nonumber \\
E_{int}^{C_i}& = & \sum_i[\sum_{j \in C_i} K_j^{cm} + \sum_{ {j,k
\in C_i}  j \le k} V_{j,k}] \label{eq:eECRA}
\end{eqnarray}
where the first sum is on partition clusters,
$K_j^{cm}$ is the kinetic energy of particle $j$ 
measured in the center of mass of the cluster containing particle $j$, 
and $V_{ij}$ is the internucleonic potential.
The algorithm that uses ``simulated annealing'' to find the most bound 
partition is known as ``Early Cluster Recognition Algorithm" ($ECRA$)
and has been extensively used in several fragmentation studies
~\cite{cherno99,dorso93,Strachan}, helping to discover that excited droplets 
beakup at an early stage.

\subsection{Results}

A snapshot sequence of a typical Ni+Ni central collision at 1500 MeV (Fig. 7) is shown.
Increasing the projectile energy, fragmentation region is reached where 
phase coexistence is found and a fragment distribution power law appears (Fig. 8).
 Critical exponent has been extracted for Ni+Ni central collision with a value $\tau = 2.18$.

\section{Conclusions}\label{CP}

Table 1 shows the critical exponent values extracted in this study and compared 
with others previously reported and obtained from percolation and 2D 
Molecular Dynamics simulations with a Lennard-Jones potential \cite{BarraHI}. 
Heavy Ion Collisions Molecular Dynamics simulations as well as genetic algorithm results 
for colossal magnetoresistance and HIV to AIDS transition
indicate that critical nucleation occurs for small and complex systems, static as well 
as transient, and suggest that they belong to the same universality class.

\begin{eqnarray*}
% \begin{alignat*}{2}  \hline
\text{\bf TABLE 1 }  \\ \hline
\text{\bf Nucleation Model}& \qquad \text{\bf $\tau$ Value }  \\ \hline
\text{HIV Cellular Automaton} & \qquad \text{$2.32$ }\\ \hline
\text {Colossal Magnetoresistance} & \qquad \text{ $2.38 $ }\\  \hline
\text{3-D MD Collision Simulations} & \qquad \text{$2.18$ }\\ \hline
\text{Cubic Lattice Percolation} \cite{BarraHI} 
 & \qquad \text{$2.32 \pm 0.02$ }\\ \hline
\text{Spherical Lattice Percolation}\cite{BarraHI} & \qquad \text{$2.20 \pm 0.1$ }\\ \hline
\text{2-D Collision Simulations\cite{BarraHI}} & \qquad \text{$2.32 \pm 0.02$ }\\ \hline
%\end{alignat*} 
\end{eqnarray*}

\section{Acknowledgements}

Work supported by National Science Foundation (PHY-96-00038),
 Universidad de Buenos Aires (EX-070, Grant No. TW98,
CONICET Grant No. PIP 4436/96), Universidad Aut\'onoma
Metropolitana-Azcapotzalco (Supercomputing Lab) and 
Asociaci\'on Latinoamericana de Biomatem\'aticas.

%\newpage

\newpage

\end{document}